\documentstyle[preprint,aps,epsf]{revtex}
\tightenlines

\begin{document}

\preprint{\vbox{\hbox{KEK-CP-071}
                \hbox{KEK Preprint 97-217}
                \hbox{November 1997}}}

\title{$B$ meson decay constant from quenched Lattice QCD}

\author{JLQCD Collaboration\\
  $^1$S.~Aoki\footnote{
    present address: Max-Planck-Institut f\"ur Physik,
    F\"ohringer Ring 6, D-80805 M\u''nchen, Germany.},
  $^2$M.~Fukugita,
  $^3$ S.~Hashimoto\footnote{
    present address: Theoretical Physics Department,
    Fermi National Accelerator Laboratory, P.O. Box 500, 
    Batavia, IL 60510, USA.},
  $^{1,4}$N.~Ishizuka, $^{1,4}$Y.~Iwasaki, $^{1,4}$K.~Kanaya,
  $^5$Y.~Kuramashi, $^5$M.~Okawa, $^1$A.~Ukawa,
  $^{1,4}$T.~Yoshi\'{e}} 

\address{
  $^1$Institute of Physics, University of Tsukuba, 
  Tsukuba, Ibaraki-305, Japan \\
  $^2$Institute for Cosmic Ray Research, University of Tokyo,
  Tanashi, Tokyo 188, Japan \\
  $^3$Computing Research Center,
  High Energy Accelerator Research Organization(KEK),
  Tsukuba, Ibaraki 305, Japan \\
  $^4$Center for Computational Physics,  University of Tsukuba,
  Tsukuba, Ibaraki 305, Japan \\
  $^5$Institute of Particle and Nuclear Studies,
  High Energy Accelerator Research Organization(KEK),
  Tsukuba, Ibaraki 305, Japan \\
  }
\date{\today}
\maketitle

\begin{abstract}
  A lattice QCD calculation of the $B$ meson decay constant
  is presented.  In order to investigate the scaling violation 
  associated with the heavy quark, parallel simulations are carried out
  employing both Wilson and the $O(a)$-improved clover actions 
  for the heavy quark.  The discretization errors due to  
  the large $b$ quark mass are estimated in a systematic way  
  with the aid of the non-relativistic interpretation approach of 
  El-Khadra, Kronfeld and Mackenzie. 
  As our best value from the quenched simulations at
  $\beta$=5.9, 6.1 and 6.3  
  we obtain $f_B$=163$\pm$16 MeV and  $f_{B_s}$=175$\pm$18 MeV 
  in the continuum limit where the error 
  includes both statistical and systematic uncertainties.
\end{abstract} 


\narrowtext


The $B$ meson decay constant $f_B$ is a fundamental quantity
needed to extract the Cabibbo-Kobayashi-Maskawa matrix element
$V_{td}$ from experiments on $B^0-\bar{B}^0$ mixing.
For this reason lattice QCD calculations of $f_B$ have been 
pursued over several years, employing either
relativistic\cite{wilson,Sheikholeslami_Wohlert_85}
or non-relativistic\cite{nrqcd} (including the static\cite{static}) 
formulation for the $b$ quark.

While there are a number of advantages with the relativistic 
formulation\cite{wilson,Sheikholeslami_Wohlert_85}, its 
basic problem for calculations of $f_B$  
has lain in the difficulty of controlling systematic errors 
associated with heavy quark mass 
whose magnitude in lattice units 
exceeds unity for the $b$ quark for a typical lattice
spacing $a^{-1}\approx 2-3$~GeV accessible in current simulations.
The formalism proposed in Ref.~\cite{El-Khadra_Kronfeld_Mackenzie_97},
however, has shed a new light on this problem:
these authors have pointed out that a Wilson-type lattice quark 
action for heavy quark can be reinterpreted as a non-relativistic 
Hamiltonian for an effective heavy quark field $Q$ of the form, 
\begin{equation}
  \label{eq:Non-relativistic_Hamiltonian}
  H = \bar{Q} \left[ m_1 
    - \frac{\vec{D}^2}{2m_2}
    - \frac{i\vec{\sigma}\cdot\vec{B}}{2m_B}
    + O(1/m_Q^2) \right] Q.
\label{eq:nrqcd}
\end{equation}
where the effective heavy quark mass parameters $m_i$ ($i=1,2, B,\cdots$) 
are functions of the bare quark mass $m_Q$ and the coupling constant. 
In contrast to the continuum where these parameters are equal 
($m_1=m_2=m_B=\cdots$), they mutually differ by $O(am_Q)$ at finite 
lattice spacing, which represents $O(am_Q)$ errors of the original action 
in the framework of the effective Hamiltonian (\ref{eq:nrqcd}).
These parameters, however, are calculable in perturbation 
theory, and effects of $O(am_Q)$ errors on $f_B$ can be 
systematically analyzed. 
In particular, we observe that errors of 
$O((m_2/m_B-1)\Lambda_{QCD}/m_Q)$ for the Wilson action 
($m_B\ne m_2$) is reduced to 
$O(\alpha_s\Lambda_{QCD}/m_Q, \Lambda_{QCD}^2/m_Q^2)$
for the $O(a)$-improved clover 
action\cite{Sheikholeslami_Wohlert_85}, for which $m_B=m_2$
holds at the tree level.

In this article we report on a calculation of  
the $B$ meson decay constant in quenched lattice QCD with the relativistic 
formalism employing this ``non-relativistic interpretation''.
In order to study $O(am_Q)$ systematic errors, 
we carry out a parallel set of simulations using both  
Wilson\cite{JLQCD_heavy-light_95,JLQCD_heavy-light_96} and clover 
quark actions over a wide range of heavy quark mass and lattice spacing.

The parameters of our simulations are listed in 
Table~\ref{tab:parameters}.  The standard plaquette action is employed 
for gluons.  
For the clover coefficient we use 
the tadpole-modified\cite{Lepage_Mackenzie_93} one-loop 
value\cite{Luscher_Weisz_96} 
$c_{\rm sw}=1/u_0^3[1+0.199\alpha_V(1/a)]$ where 
$u_0=P^{1/4}$ with $P$ the average plaquette. 
The lattice size is chosen 
so that the physical size is approximately kept at $L\approx $2~fm.
Seven values of the heavy quark hopping parameter are taken to
cover the charm and bottom quark masses, and four values
for light quark in a range $0.4m_{s} - 1.4 m_{s}$ with $m_s$ 
the strange quark mass. 
The simulations have been carried out on the Fujitsu
VPP500/80 supercomputer at KEK.  

We extract the heavy-light decay constant $f_P$ 
from the correlators of the axial vector current $A_4$ and 
the pseudoscalar density $P$ given by $\langle A_4(t)P(0)\rangle$ and 
$\langle P(t)P(0)\rangle$.  
In order to reduce statistical errors of the correlators, 
which rapidly increase as $am_Q$ increases, 
we employ the smeared pseudoscalar density 
$P^S(x)=\sum_{\vec{r}}\phi(|\vec{r}|)\bar{Q}(x+r)\gamma_5q(x)$
on the gluon configurations fixed to the Coulomb gauge.
The smearing function $\phi(|\vec{r}|)$ is
obtained by measuring the wave function of the
pseudoscalar meson for each set of heavy and light quark masses. 
As a result we are able to isolate the ground state signal from
a small time separation of $t\approx 0.8$~fm.

We adopt for the heavy-light axial vector current 
$\overline{q}\gamma_\mu\gamma_5Q$ the one-loop renormalization factor 
$Z_A(am_Q)$ newly calculated with
full inclusion of the heavy quark mass dependence\cite{Aoki_etal_97}.
The calculation is available for both Wilson and clover actions,
and it confirms the results of Refs.~\cite{Kronfeld_Mertens_93,Kuramashi_97} 
made earlier for the Wilson action.
We find that effects of finite $am_Q$ are non-negligible:
with $Z_A(am_Q)$ evaluated with the coupling constant $\alpha_V(1/a)$, 
$f_B$ for the Wilson action is reduced by 5\% (at $\beta=5.9$) to 
2\% (at $\beta=6.3$) 
compared to the value obtained with the $Z_A$ factor with the mass
dependence ignored, as employed in the previous studies. 
For the clover action the finite $am_Q$ effect works in the
opposite direction with a similar magnitude.
 
We remark that the field $Q$ is 
related to the original field $\Psi$ through
\begin{equation}
  \label{eq:heavy_quark_field}
  Q = e^{am_1/2} [1 + d_1 \vec{\gamma}\cdot\vec{D}] \Psi,
\label{eq:rotation}
\end{equation}
where $d_1$ is a known function of 
$am_Q$\cite{El-Khadra_Kronfeld_Mackenzie_97}.
The KLM factor\cite{El-Khadra_Kronfeld_Mackenzie_97} $e^{am_1/2}$ is 
evaluated including the $m_Q$-dependent 
one-loop correction\cite{Aoki_etal_97}.
We ignore the $d_1 \vec{\gamma}\cdot\vec{D}$ term, since its 
corrections to $f_B$ is expected to be at most 
1--2\% due to a small value of $d_1\approx 0.1$.

A non-trivial issue in lattice studies of heavy-light mesons 
is how to  define their masses, since the pole mass 
directly measurable from meson propagators with zero spatial 
momentum suffers from large $O(am_Q)$ errors.  
A possible choice is the kinetic mass defined by an expansion 
of the energy-momentum dispersion relation of the meson, 
\begin{equation}
  \label{eq:dispersion_relation}
  E_{\rm meson}(\vec{p})=m_{\rm pole}+\frac{\vec{p}^{~2}}{2m_{\rm kin}}+
  O(\vec{p}^{~4}).
\label{eq:kinetic}
\end{equation}
The kinetic mass $m_{kin}$, however, receives corrections from 
$O(\vec{p}^{~4})$ terms in (\ref{eq:nrqcd}) which are uncontrolled 
and hence suffer from a large $O(am_Q)$ effect\cite{Kronfeld_96}.
This leads to a pathology that 
the $b$ quark mass cannot be determined
consistently from heavy-light and heavy-heavy mesons
\cite{SCRI_95,JLQCD_heavy-light_96}.

An alternative choice may be to define a ``kinetic mass'' by taking 
the pole mass 
for a meson corrected by the difference of the kinetic and pole 
masses of the heavy quark 
$m_2-m_1$ \cite{Bernard_Labrenz_Soni_94,JLQCD_heavy-light_96}, 
\begin{equation}
  \label{eq:meson_mass}
  m_{kin} \equiv  m_{pole} + (m_2-m_1)
\end{equation}
This choice is motivated by the expectation that the binding energy of 
a heavy-light meson becomes independent of the heavy quark mass in 
the non-relativistic limit and
$(m_2-m_1)$ should thus represent the difference between kinetic and
pole masses of the meson.  We find that 
the meson mass calculated in this way does not suffer from the 
pathology observed with $m_{kin}$ defined by (\ref{eq:kinetic}).  
We adopt this definition in our analyses  
using the one-loop perturbative result\cite{Aoki_etal_97} 
for $m_2-m_1$.

\begin{figure}[p]
  \begin{center}
    \epsfxsize=85mm \epsffile{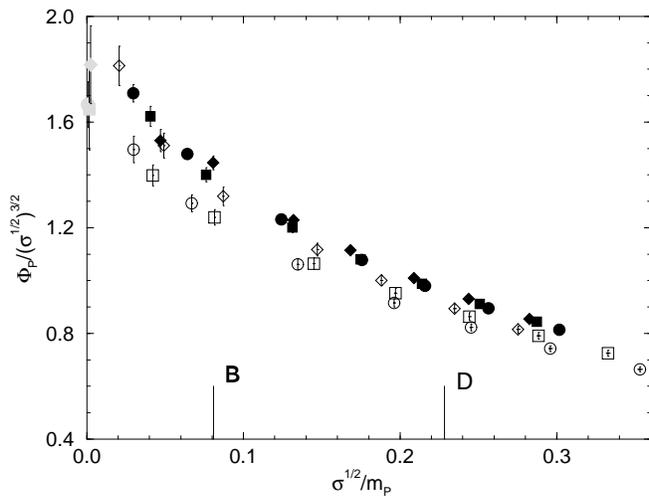}
    \caption{$\Phi_P$ as a function of $1/m_P$ 
      normalized by string tension $\sigma$.
      Filled symbols represent results with the
      clover action and open symbols with the Wilson
      action. 
      Circles, squares and diamonds correspond to results 
      at $\beta$=5.9, 6.1 and 6.3, respectively. 
      Points at $1/m_P=0$ are static 
      results\protect\cite{Duncan_et_al_95}
      at the same set of $\beta$.}
    \label{fig:heavy-light_decay_constant}
  \end{center}
\end{figure}

Let us now present our results.   
We plot  
$\Phi(m_P)=(\alpha_s(m_P)/\alpha_s(m_B))^{2/\beta_0}f_P\sqrt{m_P}$
in Fig.~\ref{fig:heavy-light_decay_constant} as a 
function of the inverse heavy-light meson mass $m_P$ for both 
 Wilson (open symbols) and 
clover(filled symbols) actions.
The light quark mass is linearly extrapolated to the chiral limit, and 
$\alpha_s(\mu)$ is calculated with the standard 2-loop definition 
where we employ the value $\Lambda_{QCD}=$ 295 MeV estimated from 
the $\alpha_V$ coupling using the plaquette average\cite{Lepage_Mackenzie_93}.

There is an ambiguity in practice as to what mass scale is to be adopted
to represent a quantity that has mass dimension.  We prefer
to use the scale that does not depend on the quark sector to facilitate
a direct comparison of the $O(am_Q)$ errors with the two different
quark actions on the common gauge configurations. Hence our natural
choice is the string tension $\sigma$, 
and the ordinate is normalized by $\sigma^{3/4}$ and the abscissa by 
$\sigma^{1/2}$ in Fig.~\ref{fig:heavy-light_decay_constant}, where 
we employ the string tension of Ref.~\cite{Bali_Schilling_92}.
Vertical lines indicate the positions of the
$B$ and $D$ mesons if one uses a phenomenological value 
$\sqrt{\sigma}=427$~MeV\cite{Cornell_potential}.
Data points plotted at $1/m_P=0$ 
are the static results\cite{Duncan_et_al_95}, to which our data 
seem to converge towards the heavy quark mass limit.
We observe that the Wilson results exhibit a small increase
as the lattice spacing decreases, 
while the clover points at three values of $\beta$  
lie almost on a single curve.  

\begin{figure}[p]
  \begin{center}
    \epsfxsize=85mm \epsffile{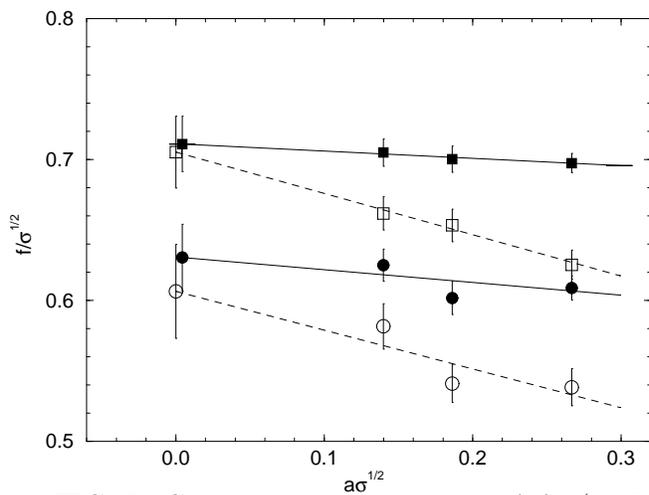}
    \caption{Continuum extrapolation of $f_B$ (circles) and
      $f_D$ (squares).
      Filled symbols represent results with the
      clover action and open symbols with the Wilson
      action. }
    \label{fig:f_B_D_continuum}
  \end{center}
\end{figure}

An improved scaling behavior with the clover action 
is more clearly seen in 
Fig.~\ref{fig:f_B_D_continuum}, where we present the continuum 
extrapolation of $f_B$ and $f_D$.
Compared to scaling violation of 11-5\% in our range
of lattice spacing $a^{-1}\approx$ 1.6--3~GeV for the Wilson case, the clover
data show a significantly reduced variation of 4--2\% over the same range
of lattice spacing.  These
magnitudes are common to $f_B$ and $f_D$.
The continuum values obtained
by a linear extrapolation agree
within the statistical error of about 5\%
between the two actions.

\begin{figure}[p]
  \begin{center}
    \epsfxsize=85mm \epsffile{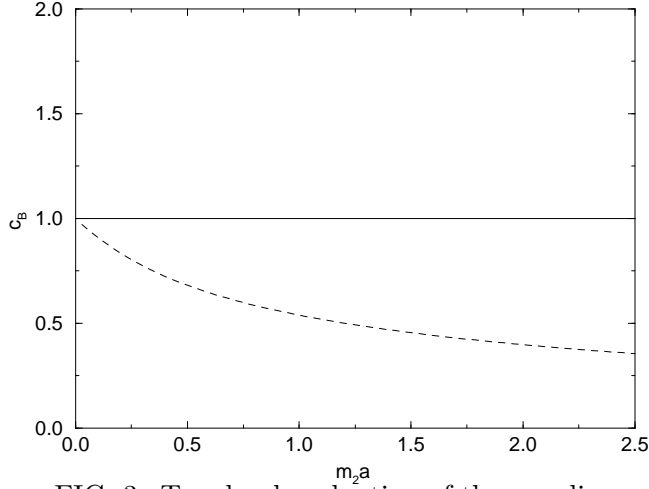}
    \caption{Tree-level evaluation of the coupling $c_B$ of
      the chromomagnetic interaction term in the
      non-relativistic effective Hamiltonian. 
      Solid and dashed lines correspond to the clover
      and Wilson actions, respectively.}
    \label{fig:c_B}
  \end{center}
\end{figure}

This agreement, however, does not necessarily mean
that systematic $O(am_Q)$ errors are all removed by the 
continuum extrapolation.
Let us discuss this point for the Wilson action for the heavy quark. 
According to the non-relativistic Hamiltonian (\ref{eq:nrqcd}), 
the size of the leading $O(am_Q)$ error in $f_B$   
is $O((c_B-1)\Lambda_{QCD}/m_Q)$ where  
$c_B\equiv m_2/m_B$.  The tree level value 
$c_B=1/(1+\sinh m_1a)$\cite{El-Khadra_Kronfeld_Mackenzie_97} 
is plotted in Fig.~\ref{fig:c_B} as a function of 
$m_2a=e^{m_1a}\sinh m_1a/(1+\sinh m_1a)$.
For $m_2a\approx 2.9-1.5$, corresponding to the $b$ quark  
at $\beta=5.9-6.3$, $|c_B-1|\approx 0.7-0.5$, and hence we expect an error 
of $O(4-3\%)$ in $f_B$ at our simulation points. 
If we linearly extrapolate $c_B$ to the continuum limit $m_2a=0$,  
$|c_B-1|$ decreases to 0.4, 
which implies an $O(3\%)$ error left unremoved.
For the $D$ meson, the value of $|c_B-1|$ is smaller 
($|c_B-1|\approx 0.4-0.3$ for the charm quark at $m_2a\approx
0.9-0.5$) and decreases faster, extrapolating to $|c_B-1|\approx 0.2$ 
at $m_2a=0$.  Thus, $O(am_Q)$ 
errors of $O(7-5\%)$ for $f_D$ at our simulation points reduces
to $O(3\%)$ in the continuum limit. 
This consideration indicates that the use of non-relativistic Hamiltonian
inherently leaves a  $m_Q$-dependent systematic error that cannot be removed 
by a linear extrapolation.
We estimate that it is of the order of 3\% for $f_B$ and $f_D$ 
in the continuum.

We need to consider
two more sources of systematic errors,
which can in principle be removed by the extrapolation procedure if
the simulation is made at high precision but in practice
are not removed from our results due to the insufficient statistics. 
One of them is $m_Q$-independent scaling violation, which is 
$O(a\Lambda_{QCD})$
for the Wilson action. 
We take the value $a\Lambda_{QCD}\approx 10\%$ at our smallest lattice spacing 
$a^{-1}\approx$3~GeV as an estimate of $O(a\Lambda_{QCD})$ scaling 
violation effects.
The other is the $O(\alpha_V^2)$
uncertainty due to the use of one-loop value for $Z_A$, 
which is $O(4\%)$ with $\alpha_V(1/a)\approx 0.2$ at $a^{-1}\approx 3$~GeV. 
Therefore, we expect a systematic error of the order of 10\% in our
results for the decay constant obtained with the continuum extrapolation.

This error analysis gives us some insight about the origin of
the scaling violation observed in Fig.~\ref{fig:f_B_D_continuum}.
We can conclude that the dominant part of the lattice spacing dependence
comes from the $m_Q$-independent $a\Lambda_{\rm QCD}$ effect, since 
the $m_Q$-dependent errors $O((c_B-1)\Lambda_{QCD}/m_Q)$ diminishes 
only little towards the continuum limit and hence contributes little to
the slope as a function of $a$.
This leads us to expect that  $f_B$ and $f_D$ exhibit a similar
slope as a function of $a$, as we indeed observe in the figure. 
The size of scaling violation actually observed  
is within a factor of two from our estimate. 

For the clover action the $m_Q$-dependent errors are reduced to 
$O(\alpha_s\Lambda_{QCD}/m_Q)$ and $O((\Lambda_{QCD}/m_Q)^2)$.  
We estimate them to be $O(1\%)$.
The scaling violation error is
$O(\alpha_sa\Lambda_{QCD})$ and $O(a^2\Lambda_{QCD}^2)$
which are of the order of 2\%.
Taking account of the $O(\alpha_V^2)$ error from $Z_A$ and that arising from 
the field rotation term $d_1\Lambda_{QCD}/m_Q\approx O(2\%)$ in 
(\ref{eq:rotation}), which is ignored in the present calculation, 
we expect systematic errors of order 5\% for the decay constant from the
clover action. The $m_Q$-independent scaling violation also dominates
the $a$ dependence of the decay constant, the contribution of
$m_Q$-dependent errors being very small.

\begin{figure}[p]
  \begin{center}
    \epsfxsize=85mm \epsffile{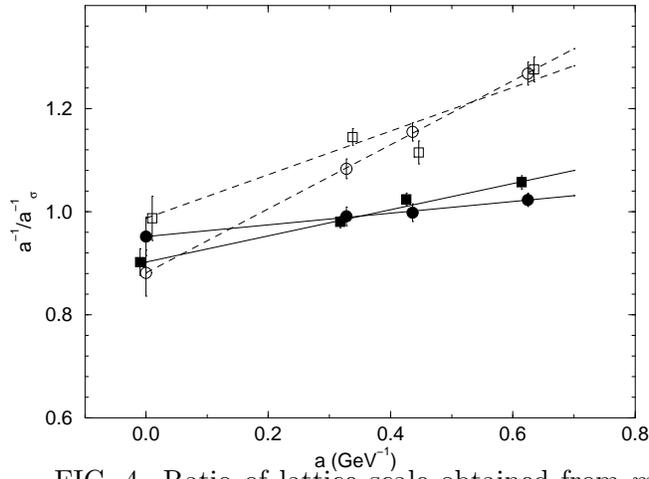}
    \caption{Ratio of lattice scale obtained from $m_{\rho}$ 
      (circles) and from $f_{\pi}$ (squares) to that from the
      string tension.
      Filled symbols represent results with the
      clover action and open symbols with the Wilson action.}
    \label{fig:lattice_scale_ratio}
  \end{center}
\end{figure}

We now examine the question of how to set the physical scale of lattice 
spacing to calculate the decay constant.
The most common in the literature is to use 
either $\rho$ meson mass $m_{\rho}$ 
or pion decay constant $f_{\pi}$ to determine the lattice scale.
In Fig.~\ref{fig:lattice_scale_ratio} we give the ratio of 
the lattice scale obtained with these quantities to 
that with the string tension.
For the clover action the $O(a)$-improved axial vector current
$A_4+c_Aa\partial_4 P$ is used to measure $f_{\pi}$ with
the one-loop value for the coefficient
$c_A$\cite{Luscher_Weisz_96}. 

As expected, the slope of the ratio is much more gentle for 
the clover action compared to that for the Wilson action.
The values in the continuum limit obtained by a linear 
extrapolation show a significant scatter, and 
the continuum limits of the ratio with the two actions 
disagree at the level of 5--10\%.
We ascribe this discrepancy mainly to smaller statistics of our Wilson 
simulation, 
and a resulting uncertainty in the continuum extrapolation. 

The continuum value of the ratio need not be 
equal to unity in the quenched approximation; the disagreement may represent
the systematic error due to quenching.  Separating the quenching error
from statistical and extrapolation uncertainties, however, is not 
possible with our present statistical accuracy.  We then take the 
dispersion of the ratio 
in Fig.~\ref{fig:lattice_scale_ratio} as 
an uncertainty of the scale
including the quenching error. We estimate it to be 10\% for the Wilson 
action and 5\% for the clover case.

We present our results for the physical value of the decay constant
in Table~\ref{tab:result}. Here  
we set the scale using the $\rho$ meson mass.  To obtain the 
ratio $f_P/m_\rho$ in the continuum limit, we combine the continuum 
values of $f_P/\sqrt{\sigma}$ and $\sqrt{\sigma}/m_\rho$ obtained 
by a linear extrapolation as given in Figs.~\ref{fig:f_B_D_continuum} 
and \ref{fig:lattice_scale_ratio}.
A direct continuum extrapolation of $f_P/m_\rho$ yields consistent
results.  
The errors quoted in the parentheses are, in the order given, statistical, 
systematic and scale errors.

We take the result from the clover action to be our
best estimate primarily because the uncertainty from scaling violation 
is smaller, but also because
our statistical ensemble is larger for this case.
Combining errors by quadrature we obtain 
 $f_B$=163$\pm$16 MeV and  $f_{B_s}$=175$\pm$18 MeV
for the B meson decay constants. For the D meson we obtain
$f_D=184\pm17$ MeV and $f_{D_s}=203\pm19$ MeV.

We have shown in this article that $B$ meson 
decay constant within a 10\% accuracy can be obtained with 
the $O(a)$-improved clover quark 
action in current lattice simulations at $a^{-1}\approx 1.6-3$~GeV.
The systematic error associated with the heavy quark is 
no longer the dominant source of uncertainty.  The 
uncertainties in the lattice scale turns out to be more important
in the present simulation.
Time-consuming full QCD simulations are perhaps indispensable to go
beyond the presently achieved accuracy in view of the fact that the
scale uncertainty involving the quenching error will be the largest
source of the uncertainty in the calculation of the heavy-light
decay constant.

\vspace*{5mm}
This work is supported by the Supercomputer 
Project (No.~97-15) of High Energy Accelerator Research Organization (KEK),
and also in part by the Grants-in-Aid of 
the Ministry of Education (Nos. 08640349, 08640350, 08640404,
09246206, 09304029, 09740226).

\newpage

\begin{table}[htbp]
  \setlength{\tabcolsep}{0.3pc}
  \begin{center}
    \caption{Simulation parameters. The lattice scale quoted is estimated 
      from $m_\rho$=770~MeV.}
    \begin{tabular}{lllll}
      action & $\beta$ & 5.9 & 6.1 & 6.3 \\
             & size & 16$^3\times$40
                    & 24$^3\times$64
                    & 32$^3\times$80 \\
       \hline
      Wilson &$N_{\rm conf}$  & 150   & 100   & 100 \\
             & $1/a$ (GeV)& 2.03(3) & 2.65(4) & 3.31(6) \\
\hline
      clover &$N_{\rm conf}$  & 540   & 200   & 166 \\
             &$c_{\rm sw}$    & 1.580 & 1.525 & 1.484 \\
             & $1/a$ (GeV)& 1.64(2) & 2.29(4) & 3.02(5) \\
    \end{tabular}
  \label{tab:parameters}
  \end{center}
\end{table}

\begin{table}[tbp]
\setlength{\tabcolsep}{0.8pc}
  \begin{center}
  \caption{Results for the decay constant in MeV unit. }
    \begin{tabular}{lll}
                        & Wilson & clover \\
      \hline
        $f_B$           & 140(11)(15)(24) & 163(9)(8)(11)\\
        $f_{B_s}$       & 159(10)(17)(27) & 175(9)(9)(13)\\
        $f_D$           & 163(13)(18)(28) & 184(9)(9)(12)\\
        $f_{D_s}$       & 180(11)(20)(31) & 203(9)(10)(14)\\
    \end{tabular}
  \label{tab:result}
  \end{center}
\vspace*{-9mm}
\end{table}

\end{document}